\pdfoutput=1

\documentclass[aps,prl,twocolumn,showpacs,amsmath,amssymb,amsfonts,long,superscriptaddress,preprintnumbers]{revtex4}
\usepackage{graphicx,color,soul}
\newcounter{mysectionnumber}
\newcommand{\mysection}[2]{\addtocounter{mysectionnumber}{1}\textbf{\arabic{mysectionnumber}.} \textsl{#1}{#2} -- }

\begin{document}

\title{A Sterile Neutrino Origin for the Upward Directed Cosmic Ray Showers Detected by ANITA}

\author{John F. Cherry}
\affiliation{Department of Physics, University of South Dakota, Vermillion, SD 57069, USA}
\email{JJ.Cherry@usd.edu}
\author{Ian M. Shoemaker}
\affiliation{Department of Physics, University of South Dakota, Vermillion, SD 57069, USA}
\email{Ian.Shoemaker@use.edu}
\date{8-23-2018}

\begin{abstract}
%We investigate the idea that the upward directed cosmic ray air shower (event 3985267) detected by ANITA originated from a sterile neutrino.  Such a neutrino could easily pass through the Earth on the steep trajectory observed for this event and is permitted to interact with $\tau$ lepton flavor through its mixing to Standard Model neutrino flavor states.  

The ANITA balloon experiment has recently observed several $\sim$ EeV cascade events at an angle below the horizon that renders any Standard Model (SM) interpretation unlikely as the Earth is significantly opaque to all SM particles at such energies. In this paper, we study a sterile neutrino interpretation of these events, calculating the angular acceptance of cascades and the relative sensitivities of several experiments to a cascade initiated by an EeV sterile neutrino.  We find that ANITA is uniquely sensitive to this type of upward directed cascade signal over a wide portion of the sky and from the direction of the two observed events has a transient acceptance roughly equivalent to that of the IceCube experiment.

\end{abstract}

\preprint{}

\maketitle

\mysection{Introduction} {.} 
The existence of the Cosmic Microwave Background (CMB) has dramatic consequences for the propagation of ultrahigh energy cosmic rays (UHECRs). It was realized almost immediately after the discovery of the CMB that the collisions of UHECRs with CMB relic photons would suppress the CR flux at high energies (the ``GZK limit'')~\cite{Greisen:1966jv,Zatsepin:1966jv}, and form a source of ultra high-energy (UHE) neutrinos around the EeV ($10^{18}$ eV) scale~\cite{Beresinsky:1969qj} from the subsequent decay of the charged pions produced in these collisions. These neutrinos have remained undetected, perhaps until now. 

The Antarctic Impulsive Transient Antenna (ANITA) was designed to search for these UHE neutrinos by detecting the radio pulses produced as the neutrinos transit the Antarctic ice. Recently ANITA reported on the detection of several events emerging at $27.4^o$ and {$35.0^o$} below the horizon with estimated shower energies of $\sim 0.6$ EeV (event 3985267)~\cite{Gorham:2016aa} and $\sim 0.56$ EeV (event 15717147)~\cite{Gorham:2018aa}, respectively. Although consistent with the characteristics of a $\tau$ lepton cascade, the problem in interpreting these events as due to a $\nu_{\tau}$ is that the Standard Model (SM) weak interaction is sufficiently strong at these energies to render the Earth quite opaque, with a transmission probability, $P_{T} \sim 4 \times 10^{-6}$ and $P_{T} \sim 2 \times 10^{-8}$, respectively~\cite{Gorham:2016aa,Gorham:2018aa} .

Mass eigenstates associated with sterile neutrinos may experience weak interactions with suppressed cross sections due to flavor mixing, which would allow them to transit the earth with greater probability. Such states are predicted in many models of neutrino masses, and may even be hinted at by short-baseline neutrino oscillations which appear to favor eV-scale sterile neutrinos{~\cite{Aguilar:2001ty,Giunti:2010zu,Mention:2011rk,Aguilar-Arevalo:2013pmq,Ko:2016owz,Alekseev:2016llm,MiniBooNE-Collaboration:2018aa}}. Current limits on the allowed mixing angles for $e$ and $\mu$ flavors are quite strong, but leave the mixing angle $\theta_{\tau 4}$ relatively unconstrained~\cite{Collin:2016aa}.  One can estimate that the mixing angle needed to convert a SM transmission probability $P_{T}^{{\rm SM}}$ to an $\mathcal{O}(1)$ sterile neutrino transmission probability requires a mixing angle $\theta_{\tau 4}^{2} \lesssim  (\log P_{T}^{-1})^{-1}$, or $\theta_{\tau 4} \lesssim 0.2$ for the cases in question. 

We propose that {sterile neutrinos} may be the {originators} of {these events}, which raises the question of the origin of a sterile neutrino flux at the EeV scale. One possibility is the decay of very massive dark matter particles ($\sim \rm EeV$) into sterile neutrinos, similar to the decaying dark matter scenarios proposed to explain the $PeV$ neutrino flux observed by IceCube~\cite{Feldstein:2013aa,Esmaili:2013aa,Ema:2014aa,Ko:2015aa}.  Another possible mechanism involves ``sterilizing'' an initial SM neutrino flux (presumed to be the original UHE neutrino flux from cosmic rays), which was recently proposed in~\cite{Cherry:2014aa,Cherry:2016aa} via a new gauge interaction under which sterile neutrinos are charged~\cite{Babu:1991at}. Models of this sort have received renewed attention~{\cite{Hannestad:2013ana,Dasgupta:2013zpn,Bringmann:2013vra,Archidiacono:2014uo,Saviano:2014esa,Mirizzi:2014ama,Chu:2015ipa,Cherry:2016aa,Archidiacono:2016kkh,Forastieri:2017oma,Jeong:2018yts,Berryman:2018jxt,Chu:2018gxk}} as a method of reconciling the Planck limits on additional light radiative species~\cite{Ade:2015xua}, with the {numerous experimental hints of a light sterile neutrino{~\cite{Aguilar:2001ty,Giunti:2010zu,Mention:2011rk,Aguilar-Arevalo:2013pmq,Ko:2016owz,Alekseev:2016llm,MiniBooNE-Collaboration:2018aa}.} 

In this {\it Letter} we explore these possibilities in detail. In Sec.~2, we investigate the observation model of the ANITA experiment for such a $\nu_4$ initiated cascade.  In Sec. 3, we compare exposure and transient acceptance calculations for a $\nu_4$ initiated cascade.  In Sec. 4 we contrast our results with predictions for a standard model neutrino progenitor and discuss future prospects.

\mysection{ANITA Observations}{.}
A sterile neutrino propagating in the $\nu_4$ state through the earth will experience the weak interaction through mixing and potentially scatter into a SM lepton state.  This can produce a $\tau$ directly through charged current (CC) interactions with nucleons or it can produce a $\tau$ indirectly by first scattering to a $\nu_\tau$ through neutral current (NC) interaction with nucleons which subsequently re-scatters to create a $\tau$, as shown in Fig.~\ref{fig:cartoon}.  Either of these channels can create an upward directed cosmic ray cascade with the properties of the ANITA events 3985267 and 15717147.  At EeV energies $\tau$ energy loss rates within the earth are sizable due to frequent scattering with nucleons~\cite{Alvarez-Muniz:2017aa}, so much so that they will lose most of their energy prior to reaching their decay point.  The phenomenon of $\tau$ regeneration~\cite{Halzen:1998aa,Bugaev:2004aa,Bigas:2008aa} can be neglected here, as any $\tau$ which survives to create an EeV energy cascade will not have undergone a full regeneration cycle. 

\begin{figure}[t!]
\includegraphics[angle=0,width=0.45\textwidth]{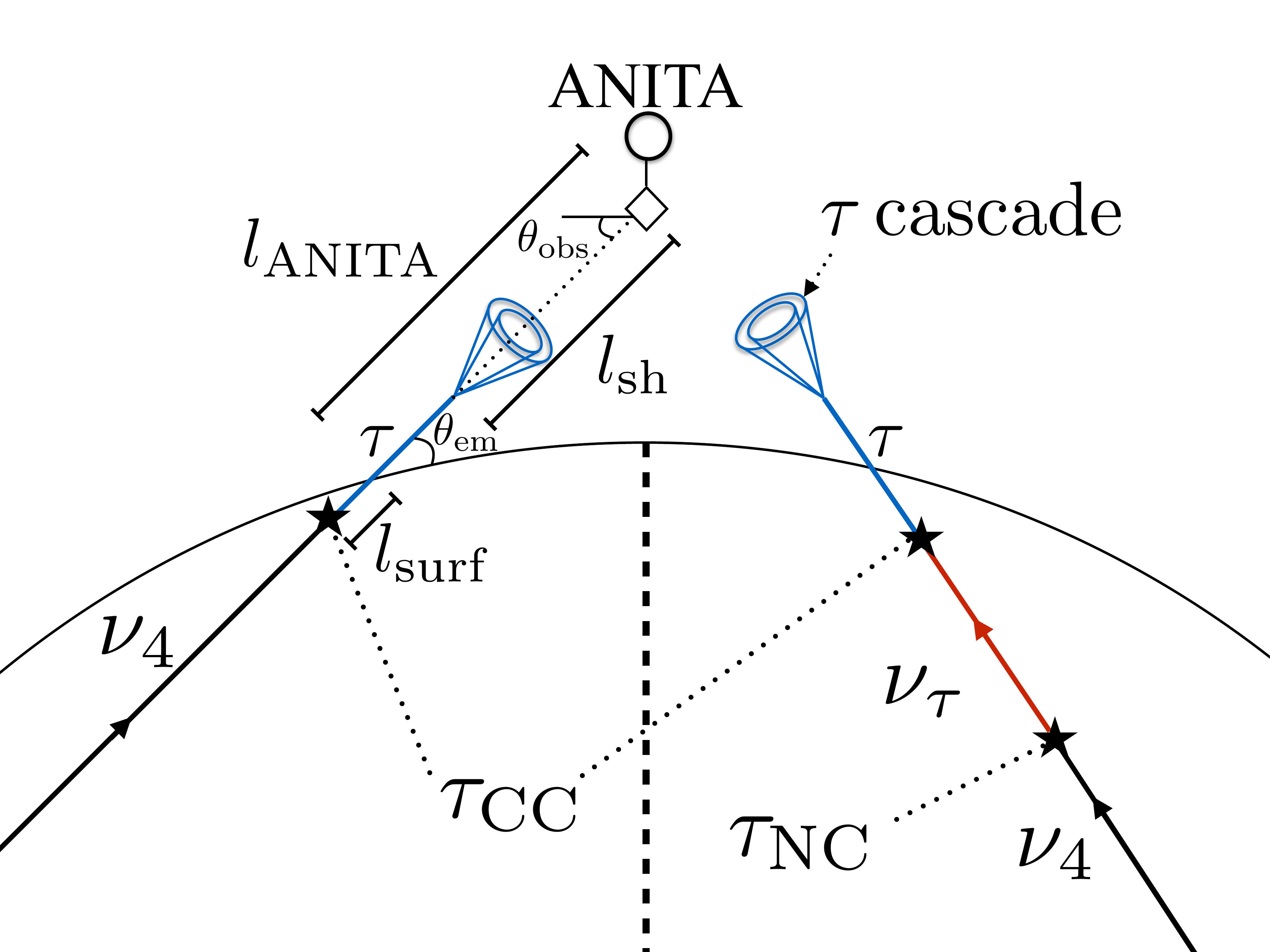}
\caption{A cartoon for the production of a sterile neutrino-induced upward pointing cosmic ray shower.  The $\nu_4$ can directly produce a shower through CC interactions (left track) or scattering into a $\nu_\tau$ flavor state via a NC interaction which then propagates and produces a cascade (right track). }
\label{fig:cartoon}
\end{figure}

For the purpose of calculating ANITA's sensitivity to $\nu_4$ initiated cascades we follow as closely as possible the collaborations own prescription for the detection of $\tau$'s created by SM neutrino interactions within the earth~\cite{Schoorlemmer:2016aa,Alvarez-Muniz:2017aa,Wolf:2017aa}, modified suitably for the propagation and interaction of $\nu_4$. The transient point source acceptance for the ANITA experiment, along a chord, $l$, at observing angle $\theta_{\rm obs}$, is given by the expression,
 \begin{equation}
 A_{i} \left( E_\nu,\theta_{\rm obs}\right) = \int_0^{l\left(\theta_{\rm obs}\right)}\frac{d P_{\rm i}\left( E_\nu ,x \right)}{dx} P_{\tau,\rm surf}\left( E_\nu ,x \right) \langle A_{\rm sh} \rangle dx\, ,
 \label{PSA}
 \end{equation}
 where x is the position coordinate along $l\left(\theta_{\rm obs}\right)$, and the index i runs over the CC and NC interaction channels.
 
A $\nu_4$ state, in the limit that the mixing with SM neutrinos is dominated by $\theta_{\tau4}$, will have a likelihood to interact within the earth's mantle of,
\begin{equation}
 \frac{d P_{\rm CC}\left( E_\nu ,x \right)}{dx} = n\left(x\right) \sin^2\theta_{\tau 4} \sigma_{\rm CC}\left( E_\nu \right) e^{-OD_{\nu_4}\left( E_\nu,x\right)}\, , 
 \label{dPintdx}
 \end{equation}
 with the nucleon number density, $n\left(x\right)$, given by the PREM density model~\cite{Dziewonski:1981:aa}, weak interaction charged current (CC) and neutral current (NC) cross sections taken from~\cite{Connolly:2011aa}, and the total transmission optical depth for a $\nu_4$ state, $OD_{\nu_4}$.   The fraction of the decay lifetime of a $\tau$ produced in the crust, $f_{\rm p}$, decreases the probability that it will reach the surface,
 \begin{equation}
 f_{\rm p}\left( E_\nu ,x \right) = \int_{0}^{l\left(\theta_{\rm obs}\right)-x} \frac{dl}{l_{\rm decay}\left( E_\nu, l \right) }\, ,
 \end{equation}
 where the instantaneous decay length along the trajectory is $c\tau_{\rm decay}\left( E_\nu, l \right)$.  To calculate the instantaneous $\tau$ decay length we take the initial $\tau$ energy to be $E_\tau = 0.8 E_\nu$ and the energy loss rate of $\tau$ leptons in Earth is taken from~\cite{Alvarez-Muniz:2017aa}.  The emergence probability that a $\tau$ created in an interaction within the earth is then simply $ P_{\tau,\rm surf}\left( E_\nu ,x \right) = e^{- f_{\rm p}\left( E_\nu ,x \right) }$.  The final emerging $\tau$ energy, $E_\tau$, is likewise computed following the ALLM model of~\cite{Alvarez-Muniz:2017aa}.  This information is then used to compute the average area of an ANITA detectable $\tau$ decay shower,
 \begin{equation}
 \langle A_{\rm sh} \rangle = \int_{0}^{l_{\rm ANITA}} \frac{d P_{\rm sh}\left( E_\tau, f_{\rm p}\right)}{dl} A_{\rm sh}\left( E_\tau, l_{\rm sh} \right) dl\, .
 \end{equation}
 Here the detectable shower area, $A_{\rm sh}\left( E_\tau, l_{\rm sh} \right)$, is calculated from the electromagnetic shower strength at distance, $l_{\rm sh}$, and ANITA detection threshold of~\cite{Wolf:2017aa}, and this is averaged with the $\tau$ decay probability, $P_{\rm sh}\left( E_\tau, f_{\rm p}\right) = e^{-\left( l/l_{
 \rm decay}\left( E_\tau \right)+ f_{\rm p }\right)}$, accounting for the fraction of a decay lifetime each $\tau$ has already spent inside the crust.  It should be noted that Equation (4.2) of~\cite{Wolf:2017aa} contains a significant, and as of this writing uncorrected, typographical error which gives the inverse of the true dependence of the electromagnetic field strength on $l_{\rm sh}$.

For the double interaction case, where a $\nu_4$ first scatters into a $\nu_\tau$ via NC interactions with nucleons, we modify Eq.~\ref{dPintdx} as follows,
\begin{equation}
 \frac{d P_{\rm NC}\left( E_\nu ,x \right)}{dx} = n\left(x\right) \sigma_{\rm CC}\left( E_\nu \right) P_{\rm \nu_4\rightarrow\nu_\tau}\left( E_\nu,x\right)\, ,
\end{equation}
where $ P_{\rm \nu_4\rightarrow\nu_\tau}\left( E_\nu,x\right) $ is the transmission probability average over histories of possible midpoints for the $\nu_4 \rightarrow \nu_\tau$ NC interaction,
\begin{equation}
 P_{\rm \nu_4\rightarrow\nu_\tau}\left( E_\nu,x\right) = \int_{0}^{x} e^{-OD_{\nu_\tau}} \frac{d P_{\rm CC}\left( E_\nu ,x^\prime \right)}{dx^\prime}\frac{\sigma_{\rm NC}}{\sigma_{\rm CC}} dx^\prime\, ,
 \end{equation}
 where $OD_{\nu_\tau}$ is the transmission optical depth of a $\nu_\tau$ of energy $E_\nu$ from the midpoint, $x^\prime$, to the end point $x$.
 
 The transient point source acceptance for ANITA is then the sum over Eq.~\ref{PSA} including both channels.  To calculate the total exposure of ANITA to $\nu_4$ initiated cascades we integrate,
 \begin{equation} 
 Exp\left( E_\nu \right) = \sum_{\rm i=CC,NC}\int\int A_{\rm i} \left( E_\nu,\theta_{\rm obs}\right) d\Omega_{\rm obs}dt\, .
  \label{EXP}
 \end{equation}
 
 %%%%%
\mysection{Exposure and Transient Acceptance}{.}
%%%%%%%%%%%%%
\begin{figure}[h]
\includegraphics[angle=0,width=0.45\textwidth]{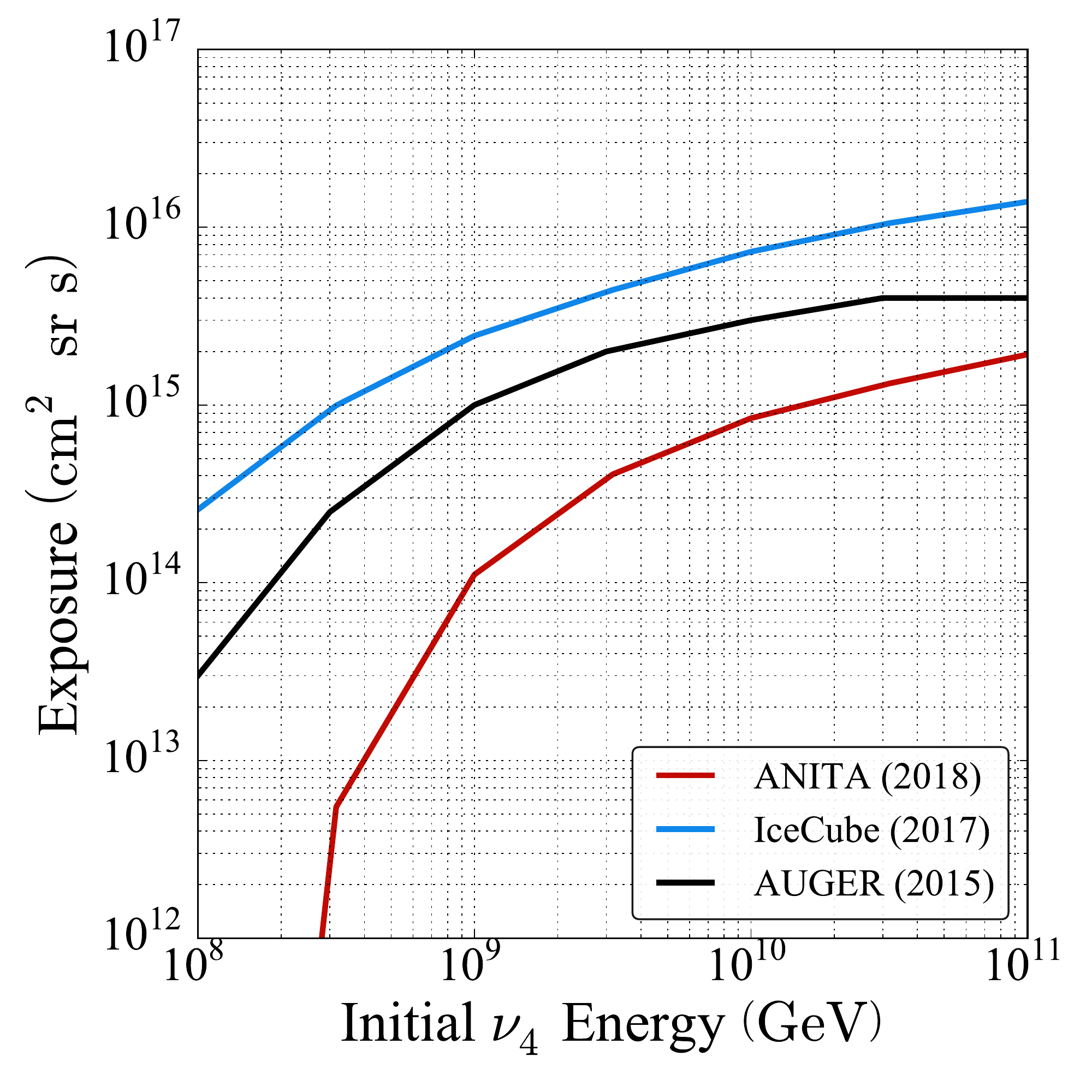}
\caption{The total isotropic exposure for the detection of upward directed $\tau$ decay cascades provided by the bulk of the earth, assuming an initial $\nu_4$ progenitor and a mixing angle of $\theta_{\tau 4} = 0.1$.}
\label{fig:exposure}
\end{figure}
%%%%%%%%%%%%%

 To gauge the the likelihood that ANITA's two anomalous cascade events may arise from $\nu_4$ initiated cascades, we compare our calculation of ANITA's exposure and transient acceptance to predictions for the IceCube neutrino observatory and the AUGER cosmic ray observatory.  Given that the AUGER experiment samples only a narrow $\sim .5^o$ band near the horizon in their search for astro-physical neutrinos~\cite{Aab:2015aa}, their exposure need only be adjusted for the reduced weak interaction cross section of $\nu_4$ in order to compare with ANITA.  IceCube, because it is capable of observing the entire sky simultaneously, has significantly increased exposure when mixing angles are sufficiently small that the Earth is transparent to $\nu_4$.  To compare with ANITA, we repeat the evaluation of Eqs.~\ref{PSA},\,\ref{EXP} for the IceCube detector making the replacement $\langle A_{\rm sh}\rangle \rightarrow \Omega_{\rm IC} l_{\rm surf}^2$, where $\Omega_{\rm IC}$ is the angular size of the IceCube detector as viewed from the point of creation of the $\tau$ lepton, and add this to the IceCube collaboration's calculation of their exposure UHE SM neutrinos~\cite{IceCube-Collaboration:2017aa} adjusted appropriately for the reduced cross section of $\nu_4$.  In what follows, we will assume a large but still allowed value of $\theta_{\tau 4} = 0.1$ for all calculations.
 
 Figure~\ref{fig:exposure} shows the total exposure of ANITA as compared to IceCube and the AUGER experiments.  Due to the ANITA balloon's limited flight time, it has the least total exposure of all experiments considered.  Compared to IceCube, ANITA has $\sim 10-20$ times less total exposure in the $\nu_4$ energy ranges which might explain the anomalous cascades they have observed.  The lack of any $\tau$ leptons observed emerging from the earth in the EeV energy range by the IceCube experiment strongly disfavors an isotropic background of $\nu_4$ as the source of ANITA's two upward directed UHECR like events.
%%%%%%%%%%%
\begin{figure}[h]
\includegraphics[angle=0,width=0.45\textwidth]{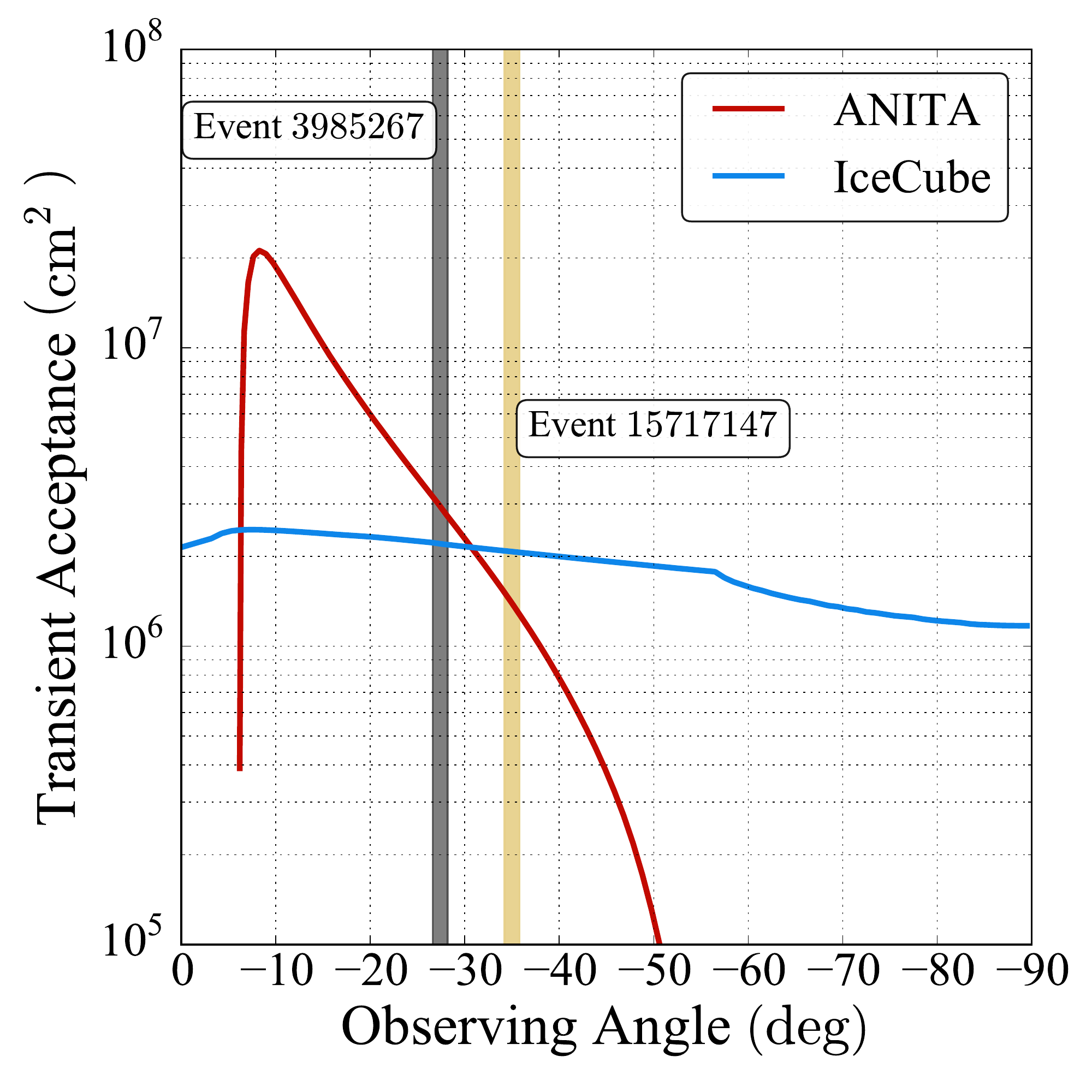}
\caption{The transient acceptance for ANITA and IceCube, assuming a mixing angle of $\theta_{\tau 4} = 0.1$ and an initial neutrino energy of $E_\nu = 1\,\rm EeV$.  Shown in gray and tan are the reconstructed trajectories of  the ANITA events.}
\label{fig:TA}
\end{figure}
%%%%%%%%%%%

Another possibility, which has been mentioned in the context of SM neutrino sources by the ANITA collaboration~\cite{Gorham:2018aa}, is that the upward directed cascades are the result of transient phenomena such as gamma-ray bursts (GRB) or supernovae (SNe).  In Figure~\ref{fig:TA} we show the transient point source acceptance of both ANITA and IceCube for a $\nu_4$ initiated cascade with $E_\nu = 1\,\rm EeV$ along side the arrival direction of the two anomalous cascades.  We find that the events fall in the portion of the sky where ANITA and IceCube have roughly equal acceptances, with detection of event 3985267 slightly favored by ANITA and event 15717147 slightly favoring IceCube.  

If we consider these two events as $\nu_4$ initiated $\tau$ lepton decay cascades we can crudely estimate the relative likelihood that either ANITA or IceCube would have observed each, assuming that the transient flux is such that the total expected number of events is one, summed over IceCube and ANITA acceptances, with an initial neutrino energy of $E_\nu = 1\, \rm EeV$.  Table~\ref{likelihood} shows the results of this estimate.  We find that at the observing angles of these two events the likelihood of ANITA observing both while IceCube detects nothing is roughly equivalent to a fair coin landing on the same side twice in a row.
\begin{table}[h]
	\centering
	\caption{Relative Likelihood}
	\label{likelihood}
	\begin{tabular}{| c |c | c |}
	\hline
	  & Event 3985267 & Event 15717147 \\
	\hline
	ANITA & $60\%$ & $40\%$ \\
	\hline
	IceCube & $40\%$ & $60\%$ \\
	\hline
	\end{tabular}	
\end{table}

%%%%%%%%%%%
\begin{figure}[h]
\includegraphics[angle=0,width=0.45\textwidth]{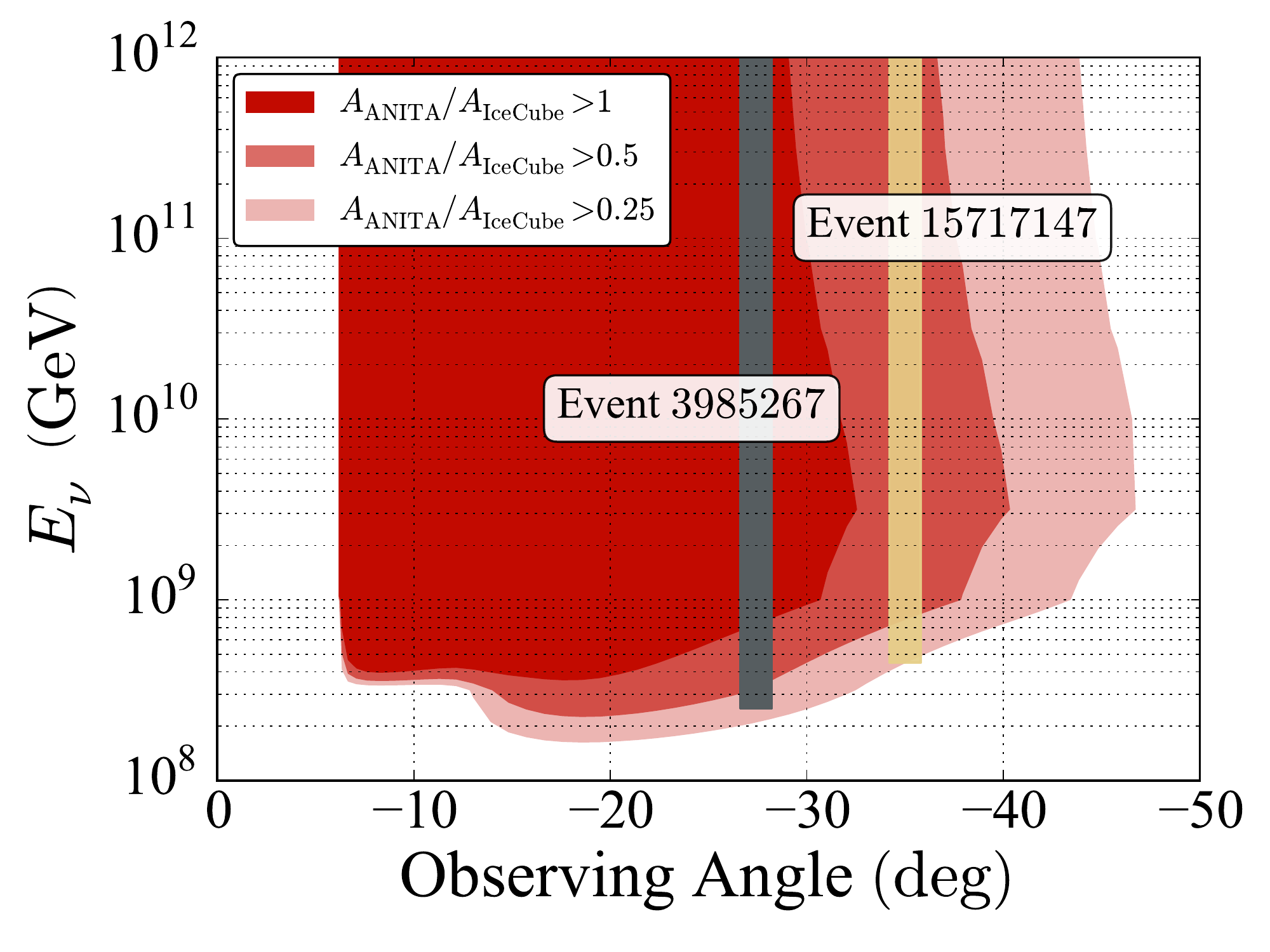}
\caption{The acceptance ratio of ANITA to IceCube, assuming an initial $\nu_4$ progenitor with a mixing angle of $\theta_{\tau 4} = 0.1$.  Shown in grey and tan are the $1\sigma$ uncertainties for the trajectories and energies of ANITA event 3985267 and event 15717147, respectively.}
\label{fig:TA_surf}
\end{figure}
%%%%%%%%%%%

In Figure~\ref{fig:TA_surf} we show the region where the ANITA experiment is comparably sensitive to IceCube for transient sources of $\nu_4$ in terms of initial neutrino energy and observing angle.  Because the cascade observation of ANITA does not put a limit on how much $\tau$ energy was lost during propagation through the earth prior to decay, the cascade energies provide only a lower bound on the initial neutrino energy.  The ANITA experiment's loss of relative sensitivity at observing angles much below $\theta_{\rm obs} \sim -40^o$ is due to several factors including angular sensitivity and alignment of the polarized antenna array~\cite{Schoorlemmer:2016aa}, and shortening of the surface distance $l_{\rm ANITA}$ which simultaneously reduces the average shower distance and increases the likelihood that a $\tau$ with high grammage will pass the experiment prior to decay.  At initial neutrino energies much less than $E_\nu < 0.3\,\rm EeV$, ANITA suffers a loss of relative sensitivity due to the reduced strength of the electromagnetic shower~\cite{Wolf:2017aa}.

%%%%
\mysection{Discussion and Conclusions}{.}
The primary motivation for considering a $\nu_4$ as the progenitor of the upward directed cascades observed by ANITA is to alleviate the need to explain the extreme fluxes of SM neutrinos needed to overcome the opacity of the earth at EeV neutrino energies.  If we consider the SM neutrino transient case, tight spatial and temporal correlation of SN2014dz~\cite{weizmann:2014dz} and the detection of event 15717147, significant at $2.7\sigma$~\cite{Gorham:2018aa}, present the interesting possibility that northern hemisphere detectors may have simultaneously observed EeV neutrino events.  Adjusting our analysis to restore neutrino cross sections and opacities to their SM values, we calculate that ANITA has a transient acceptance of {$0.15\,\rm cm^{2}$} for a $1\, \rm EeV$ SM neutrino arriving from the direction of event 15717147.  For comparison the Super-Kamiokande (SK) detector has a transient acceptance of $470\, \rm cm^2$ for a $1\, \rm EeV$ SM neutrino arriving from above the earth's horizon.  The latitude of SK, $36.2^o\, \rm N$, and declination of SN2014dz, $+38.0$~\cite{weizmann:2014dz}, yield the result that SK had SN2014dz above its local horizon $\sim 70\%$ of the time, giving a similar exposure efficiency to any SM neutrino flux from that supernova.  We can roughly estimate that if a transient flux of SM neutrinos is the source of event 15717147, SK should have experienced some {$\sim 2,200$} neutrino events in the EeV energy range during the same burst.  Estimating along similar lines for event 3985267, SK may have observed {$\sim 1$} EeV neutrino events around the same time.  SK was not running a UHE neutrino search at the time of these events, but a review of archival data would be illuminating.

The only other SM based proposal for the origin of the two upward directed cascades so far has been that of transition radiation from a UHECR neutrino interacting within the ice below ANITA~\cite{Motloch:2017aa}.  While this model requires no extension of SM physics it is simultaneously constrained by the non-observation of UHE neutrinos in the $20-30\,\rm EeV$ energy range in other detection channels and experiments.  This leads to a current prediction of $\sim 2.9\times 10^{-3}$ transition radiation events for the ANITA flight time of $52.8$ days, disfavoring two observed events in this particular model at $P \simeq 4.2\times 10^{-6}$.

Our scenario where ANITA's two upward directed showers are due to a $\nu_4$ progenitor is broadly consistent with the current set of sterile neutrino oscillation hints and constraints~\cite{MiniBooNE-Collaboration:2018aa} and also with the non-observations of transient neutrino bursts by northern hemisphere neutrino detectors.  One of the key features of this model is the kinematic decoherence of astrophysical neutrinos, which collapses them into propagating mass eigenstates prior to reaching earth.  This allows $\nu_4$ to propagate through the bulk of earth without experiencing matter suppression of their SM mixing angles~\cite{Farzan:2008aa} and allows us to neglect quantum kinetic effects~\cite{Huang:2018aa}.  There is a constraint that the $\nu_4$ state cannot be so massive that it decays back to SM particles before reaching earth after its creation.  For this we require that an EeV neutrino have a decay length of at least $1\, {\rm Gpc}$.  For the case that $\nu_4$ is a massive singlet state the lifetime is dominated by decay into 3 SM neutrinos, {$m_{4} \leq 2\,\rm MeV \times \left(.01/\sin^2 2\theta_{\tau 4}\right)^{1/6}$.}

In conclusion, our calculations lead to a road map to disentangle the mystery of ANITA's upward directed cascade events.  Because the transient $\nu_4$ acceptance of ANITA along the lines of sight of these two events is roughly equal to that of IceCube across the entire northern sky, additional transient sources of EeV energy $\nu_4$ are likely to be detected by IceCube in the future.  If the cascades are due to transient SM neutrino bursts, SK and other northern hemisphere neutrino detectors have a strong chance of observing correlated UHE neutrino events.  If the upward directed cascades continue to be detected by subsequent ANITA flights along similar trajectories with no concomitant detection in IceCube or SK, we can conclude they are likely an uncharacterized background. {We lastly note that much of the formalism employed here can be repurposed for related New Physics scenarios at ANITA, such as ``boosted dark matter~\cite{Agashe:2014yua}.''}

\begin{acknowledgments}
The authors would like to thank Benjamin Jones for originating the idea that led to this work and Alexei Smirnov and Andres Romero-Wolf for their helpful comments and feedback.
\end{acknowledgments}

\bibliography{ANITA}

\end{document}